\documentclass[useAMS,usenatbib]{mn2e}

\usepackage{times}
\usepackage{amsmath}
\usepackage{amssymb}
\usepackage{graphicx}  
\usepackage{epstopdf} 
\usepackage{hyperref}

\usepackage{color}

\newcommand{\myemail}{roskar@physik.uzh.ch}

\def\aj{AJ}
\def\araa{ARA\&A}
\def\apj{ApJ}
\def\apjl{ApJL}

\def\mnras{MNRAS}

\def\nat{Nature}

\title{Orbital Decay of Supermassive Black Hole Binaries in Clumpy Multiphase Merger Remnants}

\author[R. Ro\v{s}kar et al.]{Rok Ro\v{s}kar$^{1}$\thanks{\myemail}, 
Lucio Mayer$^1$,
Davide Fiacconi$^1$,
Stelios Kazantzidis$^{2}$,
\newauthor
Thomas R. Quinn$^3$ \&
James Wadsley$^4$
\\
$^1$Institute for Computational Science, University of Z\"{u}rich, Winterthurerstrasse 190, CH-8057 Z\"{u}rich, Switzerland\\
$^2$Section of Astrophysics, Astronomy and Mechanics, Department of Physics, University of Athens, 15784, Zografos, Athens, Greece\\
$^3$Astronomy Department, University of Washington, Box 351580, Seattle, WA 98195, USA\\
$^4$Department of Physics and Astronomy, McMaster University, Hamilton, ON, L8S 4M1, Canada\\
}

\begin{document}

\maketitle

\begin{abstract}
  We simulate an equal-mass merger of two Milky Way-size galaxy discs
  with moderate gas fractions at parsec-scale resolution including a new
  model for radiative cooling and heating in a multi-phase medium, as
  well as star formation and feedback from supernovae. The two discs
  initially have a $2.6\times10^6\mathrm{~M_{\odot}}$ supermassive
  black hole (SMBH) embedded in their centers. As the merger completes
  and the two galactic cores merge, the SMBHs form a a pair with a
  separation of a few hundred pc that gradually decays. Due to the
  stochastic nature of the system immediately following the merger,
  the orbital plane of the binary is significantly
  perturbed. Furthermore, owing to the strong starburst the gas from
  the central region is completely evacuated, requiring $\sim10$~Myr
  for a nuclear disc to rebuild. Most importantly, the clumpy nature
  of the interstellar medium has a major impact on the the dynamical
  evolution of the SMBH pair, which undergo gravitational encounters
  with massive gas clouds and stochastic torquing by both clouds and
  spiral modes in the disk. These effects combine to greatly delay the
  decay of the two SMBHs to separations of a few parsecs by nearly two
  orders of magnitude, $\sim 10^8$ yr, compared to previous work. In
  mergers of more gas-rich, clumpier galaxies at high redshift
  stochastic torques will be even more pronounced and potentially lead 
  to stronger modulation of the orbital decay. This suggests that SMBH
  pairs at separations of several tens of parsecs should be relatively
  common at any redshift.
\end{abstract}


\section{Introduction}

The relationship between the masses of supermassive black holes
(SMBHs) found at the centers of galaxies and the properties of their
hosts is one of the fundamental relations in extragalactic
astrophysics (e.g. \citealt{Magorrian:1998}; \citealt{Kormendy:1997};
\citealt{Marconi:2003}; \citealt{Haring:2004};
\citealt{Ferrarese:2000}; \citealt{Tremaine:2002};
\citealt{Gultekin:2009}; \citealt{McConnell:2012}; and references
therein).  The fact that a scaling between the two exists across
multiple orders of magnitude suggests that their growth is governed by
a common process (e.g. \citealt{Peng:2006};
  \citealt{Merloni:2010}).  Indeed SMBHs have been shown to be
ubiquitous in hosts down to disc-galaxy mass scales locally
\citep{Greene:2010}, as well as at high redshift
\citep{Schawinsky:2012}.  There is also growing evidence
  supporting the presence of SMBHs in dwarf galaxies with stellar
  masses $\lesssim10^{9}$~M$_{\odot}$ (\citealt{Reines:2011},
  \citeyear{Reines:2013}; \citealt{Koss:2014}). Combined with the
LCDM hierarchical merging paradigm of structure formation, this
implies that SMBHs should be present in a wide range of merging galaxy
pairs. Consequently, one of the channels of SMBH growth is likely to
be the coalescence of SMBH binaries.

The coalescence of the SMBH pair begins with the merging of the two
galactic cores, followed by a stage where the two SMBHs form a coupled
pair whose orbit decays due to the dynamical friction against the
stellar and gaseous background, and finally down to parsec scales due
to three-body scattering off individual stars
\citep{Begelman:1980}. Beyond the parsec scale, the decay proceeds due
to graviational wave emission \citep{Baker:2006}. The lack of observed
significant fractions of binary SMBHs or quasars implies that the
entire process from the completion of the merger to coalescence must
proceed rapidly. However, in the absence of gas, only 3-body
scattering were responsible for the decay of the binary and reaching
the gravitational-wave regime was not guaranteed due to the depletion
of scatterers from the centre as the binary hardens
\citep{Milosavljevic:2001}. Furthermore, even if the binary could
decay efficiently, the timescale would be $\gtrsim 10^9$ yrs (e.g. 
\citealt{Khan:2012, Khan:2013}).

Such a decay scenario ignores the inevitable presence of large
amounts of gas in the centers of merger remnants, which should give
rise to SMBH pairs. High-luminosity systems observed locally show an
abundance of gas near the nucleus often distributed in a disc-like
structure \citep{Sanders:1996, Davies:2004}, often with associated
star formation \citep{Davies:2004b}. It is therefore clear that to
model the decay of SMBH pairs, gas dynamics must be considered in
addition to stellar dynamical processes.

In idealized models in which an SMBH pair was embedded in a gaseous
and stellar background, \citet{Escala:2004} found that the gaseous
component provides the dominant torque for the decay of the binary,
especially when the gas distribution is in a disc
\citep{Escala:2005}. Rapid orbital decay in the presence of a nuclear
gas disc has also been confirmed in more realistic models of major
mergers, where torques during the merger efficiently funnel gas into
the central region, giving rise to a massive, dense nuclear disc
$\sim75$~pc in size \citep{Mayer:2007, Mayer:2008, Chapon:2013}. In
these models, the decay timescale is extended for stiffer gas
equations of state, but is typically found to be $<10$~Myr. The rapid
decay is primarily due to the strong dynamical friction against the
dense nuclear disc. These results are not expected to be particularly
sensitive to merger orientation \citep{Escala:2005} as long as the two
SMBHs do not depart significantly from the gas disc reforming in the
centre.

While much of the literature has focused on the rapid coalescence
problem, some recent evidence points to the existence of binary SMBHs
in systems undergoing the final stages of the merger, where the
central region is heavily obscured \citep{Fabbiano:2011}. The
implication of these observations is that binary SMBHs may be common
in luminous merger remnants, but were not previously observed due to
obscuration. In this case, the process of orbital decay from 100~pc
scale to 1~pc scale may actually take longer than previously
considered.

If the ISM in which the SMBH pair is embedded is highly inhomogeneous,
as is expected in the nuclear regions of galaxies
\citep{Wada:2001, Wada:2001b}, the torques exerted on the SMBH may be
reduced. Nevertheless, \citet{Escala:2005} found that the range of
decay timescales increased by only factor of 3 for a reasonable span
of ISM clumping factors. Recently, \citet{Fiacconi:2013} modeled the
orbital decay process in hydrodynamic simulations of SMBH pairs
orbiting in a nuclear disc. Their nuclear discs were initialized to
include clumps, whose mass spectrum is consistent with the mass
spectrum of dense gaseous clouds typically found in galaxy centers. 
With the inclusion
of disc inhomogeneities the SMBH binary orbit decay in these models
becomes highly stochastic, depending on the gravitational interactions
of the individual SMBHs with the clumps, which in many cases approach
or even exceed the SMBH mass. The interactions between clumps and the
SMBHs may accelerate or delay the decay, depending on the conditions
of the interaction. In some cases, the resulting decay timescales were
extended by as much as an order of magnitude. Their results highlight
the fact that an inhomogenenous, multi-phase medium may play an
important role in the sequence of events that eventually lead to the
SMBH coalescence.

In this paper, we push this exploration further by attempting to
model an inhomogeneous, multiphase ISM in a hydrodynamic simulation of
an equal-mass merger. The initial conditions of our model are very
similar to those of used in \citet{Mayer:2007}, but unlike in this
previous work we move away from an idealized modeling of the ISM based
on an effective equation of state, and instead allow the gas to cool,
form stars, and be affected by stellar feedback. These processes
profoundly affect the nature of the central region immediately after
the merger, qualitatively changing the orbital decay process.

This paper is organized as follows: in Sect.~\ref{sec:methods} we
first briefly summarize the salient points regarding the simulation
code \textsc{Gasoline}, and then discuss in detail our modifications
of the code; in Sect.~\ref{sec:resampling} we describe the particle
resampling method we use to increase the effective resolution of the
simulation; in Sect.~\ref{sec:morphology} we describe the
morphological evolution of the merger remnant; in
Sect.~\ref{sec:orbit_decay} we discuss the effect of the
multiphase medium on the SMBH binary orbit evolution; we conclude in
Sect.~\ref{sec:conclusions}.

\section{Methods}
\label{sec:methods}

\subsection{Simulation Code}

Our simulations are run using the Smooth Particle Hydrodynamics (SPH)
code \textsc{gasoline} \citep{Wadsley:2004}, which is an extension of
the $N$-body treecode \textsc{Pkdgrav} \citep{Stadel:2001}. All
analysis was performed using the open source \textsc{pynbody} package
\citep{Pontzen:2013} and the \textsc{IPython} environment
\citep{Perez:2007}. The simulation uses standard prescriptions for
star formation from \citet{Stinson:2006} using a Salpeter IMF. Star
particles are spawned from a gas particle whose density $\rho >
\rho_{\mathrm{thresh}}$ and $T < T_{\mathrm{thresh}}$, where
$\rho_{\mathrm{thresh}}$ and $T_{\mathrm{thresh}}$ are density and
temperature thresholds chosen such that the stars are forming in the
densest, coldest gas regions found in the simulation. Due to the large
dynamic range spanned by the simulation from the initial galaxy-scale
stages to the final stages where we focus only on the inner regions,
we have had to adjust these star formation parameters at different
times. Initially, when the two merging systems are still distinct, we
use a star formation density threshold $\rho_{\mathrm{thresh}}=0.1
\mathrm{~amu~cm^{-3}}$ and a temperature threshold
$T_{\mathrm{thresh}} = 1.5\times10^4 \mathrm{~K}$. However, as the
simulation progresses and the gas phase includes low-temperature,
high-density material, we revise this prescription to
$\rho_{\mathrm{thresh}} = 10^4 \mathrm{~amu~cm^{-3}}$ and
$T_{\mathrm{thresh}} = 200 \mathrm{~K}$. We discuss the multi-phase
simulation technique and particle splitting in
Sect.~\ref{sec:resampling} below. Due to short timescales in the
galactic centre, we also adjust the star formation timestep to
$10^4$~yr (for typical galaxy-formation simulations it is $10^6$~yr).

The supernova feedback is the ``blastwave'' feedback from
\citet{Stinson:2006}, which attempts to model the expansion of
supernova-driven bubbles by mimicking the ballistic phase of the shock
triggered by the supernova explosion. To mimic this phase, cooling is
turned off for the timescale of the snowplow phase of the shock,
calculated based on the instantaneous SN energy input and the ambient
density. Such a feedback prescription overcomes the difficulty in
distributing SN energy radiatively in the ISM; due to the implicit
optically thin gas modelling and the high densities involved, the
cooling timescales are extremely short. The blastwave feedback allows
us to efficiently couple the SN energy to the ISM. In addition, the
feedback model pollutes the gas with metals produced in SN Ia, SN II,
and AGB stars.

In the simulations presented in this paper, we make use of the
standard atomic hydrogen cooling function, which has a temperature
floor at $\sim10^4$~K. We ignore contributions to the cooling function
from metals above $10^4$ K, which could have an impact on returning
the gas expelled via feedback back to the central region. However,
even with metal-line cooling, we expect the cooling timescales for
$10^6$~K coronal gas to be much longer than the SMBH coalescence
timescale (of order a few million years based on \citealt{Mayer:2007}
and \citealt{Chapon:2013}). Therefore, while not including the
metal-dependent cooling may certainly influence the long-term
post-merger evolution, we don't expect for it to significantly alter
the properties of the rather short nuclear disk rebuilding phase
covered in this paper.  We allow the gas to cool below $10^{4}$~K
including metal lines using the empirical fit from
\citet{Maschenko:2008} based on calculations by \citet{Bromm:2001}
until the gas is optically-thin to stellar radiation, and then switch
to a novel thermal balance prescription for high density optically
thick gas, as described in the next section.

\subsection{Thermodynamics of the high density gas phase}

The gas equation of state plays a decisive role in the fate of
supermassive black hole binaries \citep{Mayer:2007, Mayer:2008,
  Chapon:2013}. In particular, the behaviour of the cold, high
density, optically thick gas phase was poorly modeled in previous
simulations adopting a prescribed equation of state.  Here, we improve
on previous work by including star formation in the coldest, densest
gas, as well as stellar feedback, during all stages of the
simulation. In addition, we implement a table for equilibrium
temperatures based on \citet{Spaans:2000} appropriate for $\rho_{\rm
  gas} > 0.1$~amu~cm$^{-3}$. The table gives an equilibrium
temperature given a gas density following a detailed calculation
including the relevant radiative processes in the densest gas
phase. The model has been calibrated using 2D radiative transfer
calculations for irradiated clouds in starburst environments
\citet{Spaans:2000}. These include stellar UV heating on dust and IR
dust emission, photoelectric heating effect on dust, cosmic ray
heating trapping of molecular and atomic lines in presence of a
photodissociation layer and local turbulent velocity dispersion. The
model is essentially an upgraded version of that adopted in
\citealt{Klessen:2007}.  It assumes a star formation rate of 100
M$_{\odot}$~$\mathrm{yr^{-1}}$ when computing the UV flux from stars
and ionization equilibrium between species.

\begin{figure}
\centering
\includegraphics[width=\columnwidth]{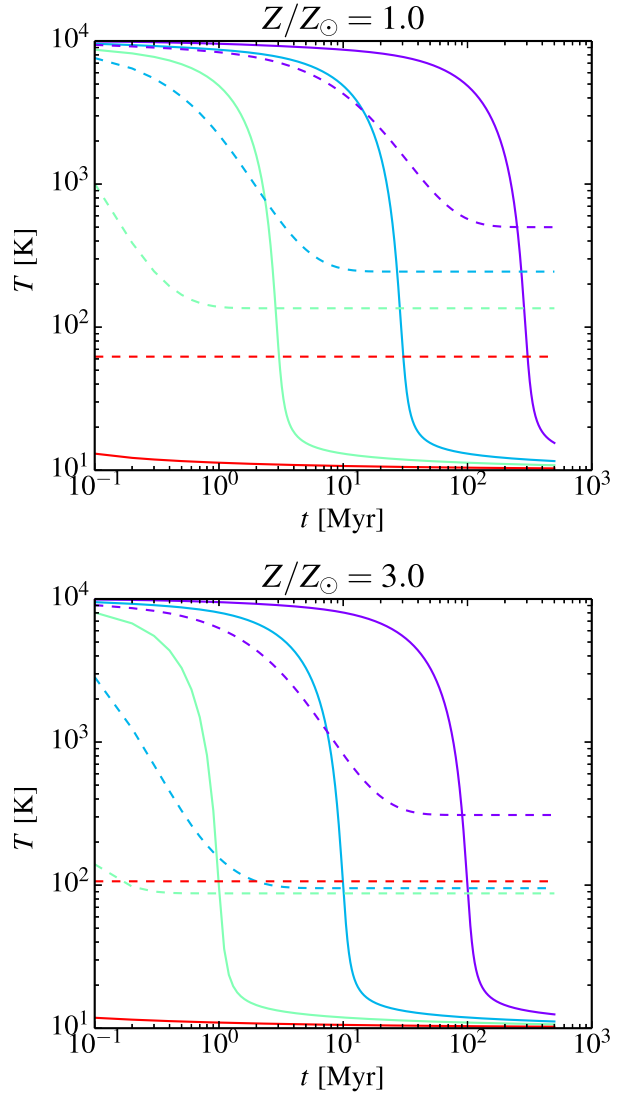}
\caption{Temperature evolution of a single particle using the standard
  low-temperature cooling curve (solid lines) and our temperature
  correction using calculations from \citet{Spaans:2000} (dashed
  lines). Purple, turquoise, green and red colours represent densities of
  1, 10, 100 and $10^4$~amu~cm$^{-3}$ respectively.}
\label{fig:cooling}
\end{figure}

In practice, we implement this cooling table as a temperature
correction on top of the usual temperature calculation. If the
particle's density is in the range described by the table, then we
modify its temperature $T_{\rm p}$ by a factor $\Delta T = (T_{\rm eq} - T_{\rm
  p})/T_{\rm eq}$, where $T_{\rm eq}$ is the equilibrium temperature
interpolated from the cooling table. In this way, the particles are
pushed toward the equilibrium solution rather than simply assigned a
temperature.  Fig.~\ref{fig:cooling} shows the temperature evolution
of a single particle at different densities, for solar (top panel) and
super-solar metallicity (bottom panel). The dashed lines show the
cooling trajectories using our temperature correction, while the solid
lines show the standard low temperature cooling. One can see clearly
that the temperature correction initially accelerates the cooling, but as
the gas gets colder it introduces a higher temperature floor at a given density.
Essentially, this allows us to capture the reduced efficiency of
cooling in regions of high optical depth in the very inner, dense region
of our system, without having to resort to a full radiative transfer
calculation. Note that the temperature correction is applied after the
energy calculation, so as to make sure that the cooling still takes
place at a reasonable rate.

\begin{figure}
\centering
\includegraphics[width=\columnwidth]{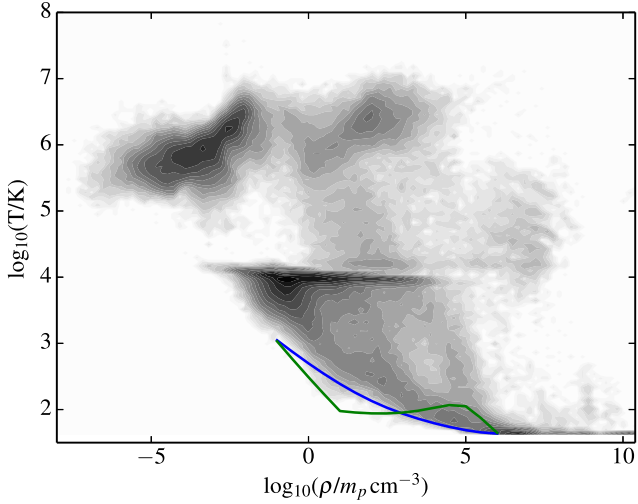}
\caption{Phase diagram for the gas during the final stages of the
  simulation. The blue and green lines delineate the gas equilibrium
  from \citet{Spaans:2000} for low ($Z < Z_{\odot}$) and high ($Z >
  Z_{\odot}$) metallicities respectively.}
\label{fig:rhoT}
\end{figure}

In Fig.~\ref{fig:rhoT} we show the phase diagram using this low
temperature cooling correction. We chose an output at the moment of
the final apocentre, when the softening is reduced to 1~pc to ensure
that the high-density end of the distribution is well-populated. The
blue and green lines show the modified cooling table values for
$Z_{\odot}$ and $3 Z_{\odot}$ respectively.

\begin{figure}
\centering
\includegraphics[width=\columnwidth]{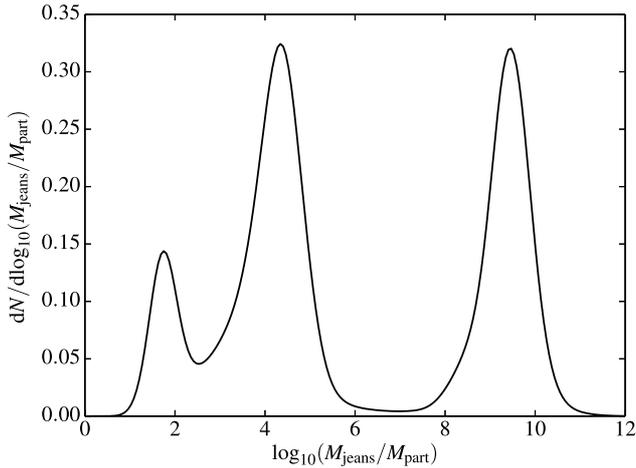}
\caption{Ratio of particle mass to local jeans mass during the final
  stage of the simulation. Our imposed pressure floor ensures that the
  local jeans mass $M_{\mathrm{jeans}}$ is always resolved by
  approximately 10 particles, ensuring that any clumping we see in the
  simulation is physical.
  }
\label{fig:jeans}
\end{figure}

Since we are concerned with modeling the inhomogeneous ISM and we are
allowing the gas to cool to low temperatures, we must take care to
guarantee that any fragmentation remains physical. We ensure that the
gas particle's implied Jeans scale is resolved at a given temperature
and density by imposing a pressure floor constraint following
\citet{Agertz:2009}. The minimum pressure is set to $P_{\rm min} =
\epsilon G h^2 \rho^2$, where $\epsilon = 3.0$ is a safety factor, $G$
is the gravitational constant, $h$ is the smoothing length and $\rho$
is the particle density. In Fig.~\ref{fig:jeans} we show the
distribution of the ratios of $M_{\mathrm{jeans}} / M_{\mathrm{part}}$
during the final part of the simulation when the force resolution is
1~pc, showing that the Jeans mass is resolved by $\gtrsim 10$
particles everywhere in the simulation are resolved and therefore any
fragmentation and clumping we see is physical.

\subsection{Particle Resampling during the Galaxy Merger}
\label{sec:resampling}

The system is initialized as a merger between two Milky Way-like discs
each with a SMBH particle embedded in the centre. The same initial
conditions were used as a starting point for the simulations presented
in \citet{Mayer:2007}. The initial systems are equilibrium discs
sampled with $10^5$ gas particles and $2\times10^5$ star particles
($10^5$ particles in the disc and the same number in the bulge)
embedded in dark matter halos with $10^6$ particles. The particle
masses are $4\times10^4$~M$_{\odot}$ for gas and
$8\times10^4$~M$_{\odot}$ for star and dark matter particles. The
masses of the three components in each of the initial discs are
$4\times10^9 \mathrm{M_{\odot}}$, $4\times10^{10}\mathrm{M_{\odot}}$,
and $1\times10^{12} \mathrm{M_{\odot}}$ for gas, stars, and dark matter
respectively.  With these choices the resulting galaxy model is a
typical massive late-type spiral, consistent with the predictions of
abundance matching at $z=0$ (e.g. \citealt{Behroozi:2013}) and having
a moderate but typical gas fraction in the disc for Sb/Sc galaxies of
10\%. Both galaxies host a SMBH modeled as a collisionless particle of
$2.6\times10^6$~M$_{\odot}$ embedded in a bulge of
$8\times10^9$~M$_{\odot}$. The initial softening lengths are 100~pc
for all particle species. Note that we use several species of dark matter
particles in order to increase the effective resolution in the inner
regions. We have verified that the individual discs are in acceptable
equilibrium by evolving them in isolation for several Gyr. The centres
of the two halos are initially separated by $>500$~kpc and the two
systems are on an in-plane parabolic orbit with a pericenter of
50~kpc, appropriate for a cosmologically-motivated merger
geometry. Star formation and feedback are turned on from the beginning
in order to provide a multi-phase gas medium and a variety of stellar
populations, which are crucial for the modelling of the state of the
central region just after the merger.

We stop the simulation just before the second passage (at
approximately 4.9 Gyr), when the two cores are separated by $\sim
5$~kpc. We define a spherical region of 35~kpc around the merger
remnant for particle resampling.  This is sufficient to ensure that
the boundaries will not interfere with the centre on timescales of
interest ($\sim 100$~Myr). We split these ``parent'' particles into
eight ``child'' particles and distribute them randomly according to
the SPH smoothing kernel around each parent. The child particles
receive $1/8$ of the mass of their parents and inherit the same
velocity, ensuring that we conserve mass and angular momentum in the
system. We compute the densities of the child particles according to
the kernel, but we distribute other simulation quantities
(temperature, metals, feedback information etc.)  to the child
particles via the standard SPH scatter scheme. For this final step, we
restrict ourselves to computing the gather radius using only 16
neighbors in order to prevent excessive blurring of boundaries in the
flow. Using this technique we attempt to minimize the introduction of
random noise that would result from a more crude resampling of the
multiphase medium. The softenings at this stage are set to 50 pc for
the stars and gas. The resampled system includes $7.3\times10^5$ gas,
$3.8\times10^6$ star, and $2.5\times10^5$ dark matter particles. Note
that like in \citet{Mayer:2007}, we do not split the dark matter
particles, although we do split the star particles. The particle
splitting is accomplished in part by reusing routines from
\textsc{skid}\footnote{\url{https://hpcforge.org/projects/skid/}}
\citep{Stadel:2001} to compute the densities and neighbor lists for
the gather-scatter scheme. 

\begin{figure*}
\centering
\includegraphics[width=\textwidth]{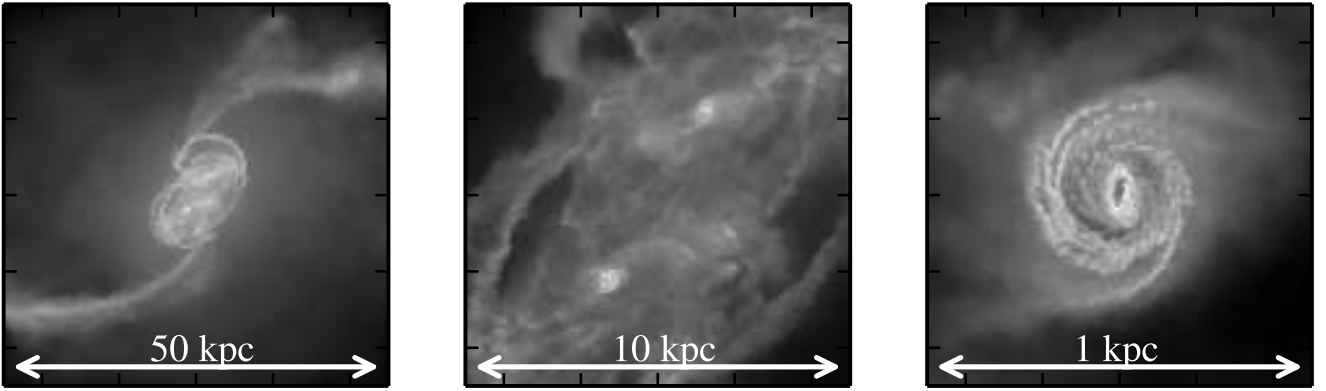}
\caption{Maps of the gas density of the system just before the final
  stage of the simulation. The two cores are clearly visible and the
  SMBHs are embedded in their centers. }
\label{fig:ic_zoomin}
\end{figure*}

Fig.~\ref{fig:ic_zoomin} shows the system on several
different scales just after the splitting.  The left panel shows the
resampled region.  The two nuclear disc cores are easily identifiable
in the middle panel, which also clearly shows the turbulent nature of
the gas. The rightmost panel zooms in on the central region, showing
the dense disc around one of the SMBHs about 500~pc across.

During the initial stages of the merger, the SMBHs stay
within a softening length of the potential minimum of the parent
disc. However, when we move to higher resolution, this displacement
could have an effect on the efficiency of the pairing, since we are
also changing the force resolution. In order to prevent the SMBHs from
leaving the centers of their respective host cores due to this change
in resolution, we move the SMBHs to match the centre of mass and
velocity of their respective discs. For this recentering, we determine
the center of mass and velocity of particles within a 1~kpc sphere
around each SMBH.

A final modification of the simulation is done at the last apocentre
passage before the completion of the merger. At this point we increase
the force resolution to 1 pc for the gas particles and the SMBH
particles in order to allow for the possibility of the SMBH orbits
decaying to the parsec scale, in a similar fashion as in
\citet{Mayer:2007} and \citet{Escala:2004}. The stars which have
already formed keep their softening, but newly formed stars have the
softening of their parent gas particles. This results in a mismatch
between the mass and force resolution in the stellar
component. However, we expect the most important contribution to
dynamical friction to come from particles closest to the SMBHs,
i.e. those in the nuclear discs. This means that for the purposes of
the orbital evolution, the newly formed stars and the gas particles
will have the largest effect and those both have resolution of 1
pc. Just as in the particle splitting step, we recentre the SMBHs with
respect to the centre of mass and bulk velocity of their surrounding
gas. After this final refinement, the simulation is evolved until the
SMBH orbit decays down to the scale of the softening length.

\section{Results}

\subsection{Post-Merger Phase}
\label{sec:morphology}

\begin{figure}
\centering
\includegraphics[width=\columnwidth]{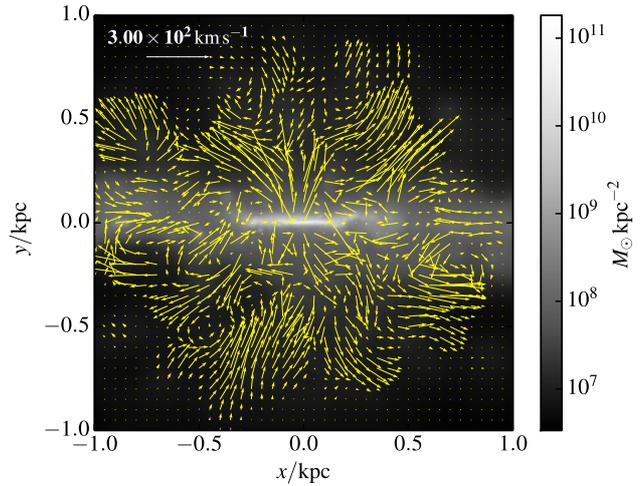}
\caption{Density map shortly after the merger with velocity
  overlaid. The velocity field is generated for a thin slice, while
  the map is showing a projected column density. Supernova-powered
  outflows from the central region with velocities of several hundred
  km/s are not uncommon.}
\label{fig:outflows}
\end{figure}

During the final passages before the completion of the merger, the two
cores undergo substantial starbursts breaking up the homogeneity of
the ISM. These startbursts, reaching star formation rates of $\sim 80
\mathrm{~M_{\odot} \, yr^{-1}}$, have the effect of blowing out the
gas from the central region, delaying the formation of the nuclear
disc for the final SMBH orbit decay. Fig.~\ref{fig:outflows} shows an
edge-on map with the velocity field overlaid just before the final
pericentre. The supernova-powered winds stir the gas surrounding the
central disc, causing asymmetric outflows and inflows, as is cleary
seen below the plane. We consider the merger to be complete when the
two cores fully merge, which occurs at time $t=5001$~Myr since the
start of the simulation. In the following analysis, we denote times in
terms of the time elapsed since the completion of the merger in units
of the orbital time, $\tau$, measured at 100~pc.

\begin{figure*}
\centering
\includegraphics[width=\textwidth]{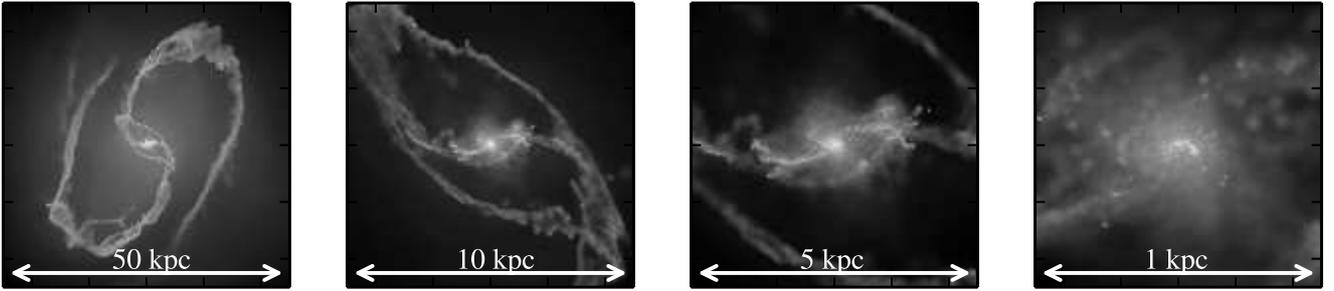}
\caption{Sequence of face-on projections of the gas distribution at
  $t=5003$~Myrs ($\tau \sim 10$, where $\tau$ is time elapsed in units
  of orbital time at 100~pc), just after the two cores become fully
  merged. The tidal tails are broken up due to feedback and multiphase
  ISM, reducing the efficiency of accretion into the central region.}
\label{fig:zoomin}
\end{figure*}

\begin{figure*}
\centering
\includegraphics[width=\textwidth]{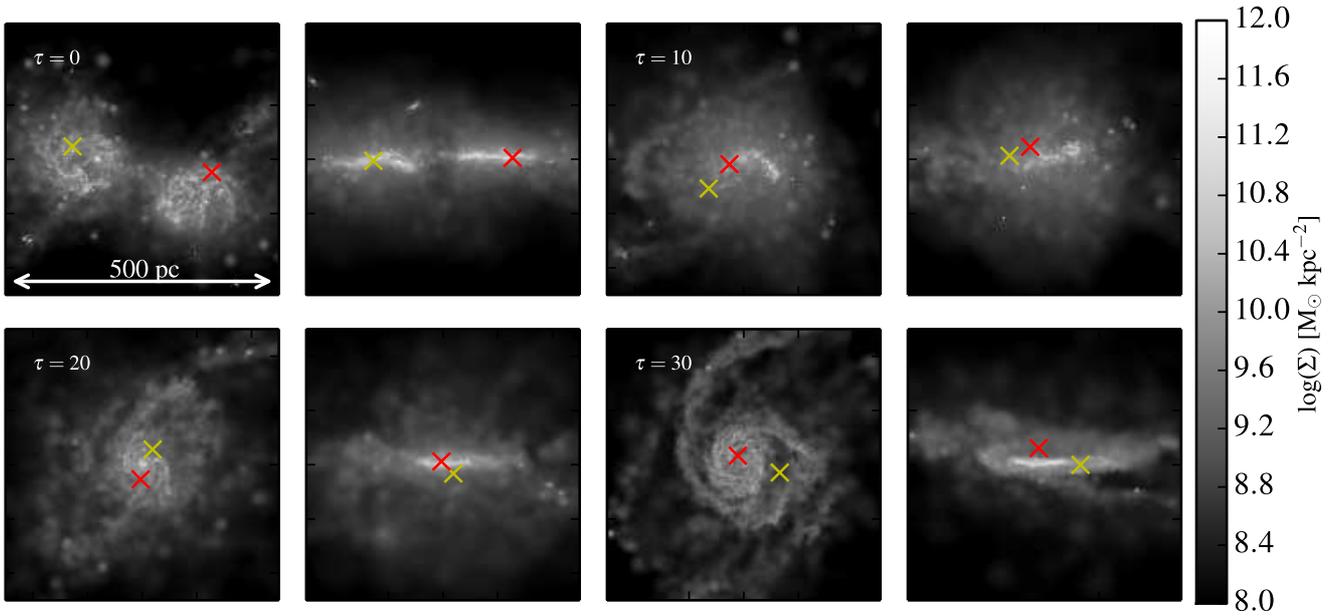}
\caption{Face-on and edge-on gas density projections at several times
  during the final phase of the merger, after the last apocentric
  passage. The two SMBH particles are indicated by yellow crosses.}
\label{fig:filmstrip}
\end{figure*}

In Fig.~\ref{fig:zoomin}, we show a single output at several different
scales, just after the two cores have fully merged. On the scales of
tidal tails (several kpc), we can see that the gas structure is not
smooth but has instead become clumpy and disordered due to the
multiphase nature of the ISM. In the inner region, the gas structure
is highly irregular at this stage, without any clear evidence of
ordered bulk motion. Furthermore, although the gas is funneled into
the centre the star formation rate remains at several
M$_{\odot}$~yr$^{-1}$ and within several $10^7$~yrs much of the gas
that makes it to the centre is depleted, as can be seen in the bottom
two panels of Fig.~\ref{fig:filmstrip}, due to feedback and star
formation.


\begin{figure}
\includegraphics[width=\columnwidth]{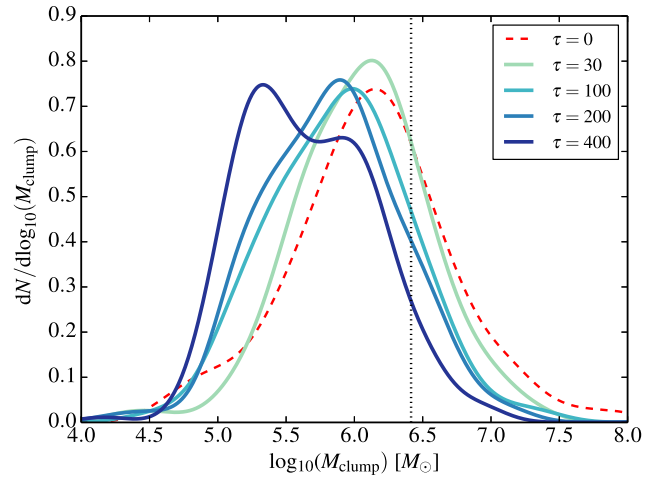}
\caption{Clump mass distributions at four different times. The vertical
  line marks the mass of the SMBHs.  }
\label{fig:clump_mass_histogram}
\end{figure}

The violent nature of the final encounters breaks the in-plane
symmetry of the merger, manifested in the misalignment of one of the
discs with respect to the $xy$ plane as can be seen in the top right
panel of Fig.~\ref{fig:filmstrip}.  The left and right column pairs
of the figure show face-on and edge-on projections respectively at
different times.  Although we took care to set the SMBH into the
centre of mass and velocity of their respective cores at the final
resampling, it is clear that they are easily tossed out of the
plane. This most likely results from interactions with the massive
clumps, which in many cases exceed the mass of the SMBHs. To quantify
the importance of dense clumps, we use the group finder
\textsc{skid}
\citep{Stadel:2001} to obtain groups of gravitationally bound
particles in every output. We set a mean density threshold to
$\rho_{\rm min} = 2~m_p \, \mathrm{cm^{-3}}$ and a linking length for
the Friends-of-Friends algorithm to five times the minimum softening
length, i.e. $\tau_{FF} = 5\mathrm{~pc}$. Note that we use both, gas
and stars in clump identification and $\rho_{\rm min}$ pertains to
density calculated from both particle species. The density threshold
ensures that we only identify the dynamically most interesting,
densest particle groups. We show the resulting clump mass
distributions at several representative times during the system's
evolution in Fig.~\ref{fig:clump_mass_histogram}. The vertical
dashed line indicates the mass of the SMBH particles, which some of
the clumps clearly exceed, in particular immediately after the
merger. At that time, many of the clumps are quite gas-rich, but due
to the high densities they quickly convert much of their mass to
stars. The clumps therefore quickly become akin to dense star clusters
rather than gas clumps. We discuss the impact of these massive clumps
on the SMBH orbital decay in Sect.~\ref{sec:orbit_decay}.

\subsection{Disc Rebuilding Phase}

\begin{figure}
\centering
\includegraphics[width=\columnwidth]{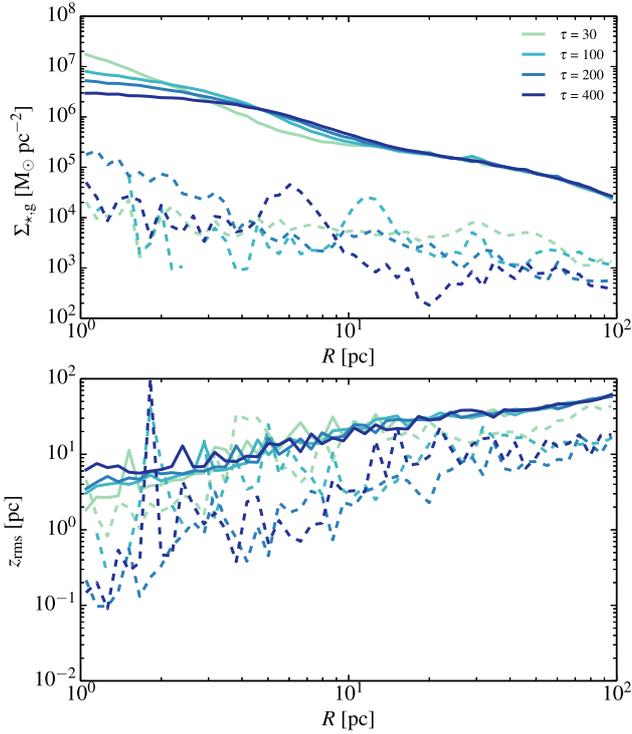}
\caption{Density and $z_{\rm rms}$
  as a function of radius (top and bottom panels respectively) for stars
  (solid) and gas (dashed) at four representative times, as indicated.
  }
\label{fig:density_profiles}
\end{figure}

The central region quickly (in a few Myrs) recovers from the starburst
and the circumnuclear disc begins to reform.
Fig.~\ref{fig:density_profiles} shows the time evolution of 
stellar and gas density profiles in the central 100~pc (top panel) and the
radial dependence of $z_{\rm rms}= \frac{1}{N}\sqrt{\sum^N_{i=0} z_i^2}$
(non-parametric proxy for thickness; bottom panel) at different times.
Owing to a period of intense star formation during the
merger, the stellar density here exceeds the gas density by a factor
of 100-1000. This is very different than the situation explored in
previous studies where the nuclear gas disc dominated the mass
distribution (e.g. \citealt{Mayer:2007}). Furthermore, while a nuclear
disc does form, the accretion is hindered by the fact that much of the
mass is locked up in massive clumps. As can be seen from
Fig.~\ref{fig:clump_mass_histogram}, the clumps form even at late
times (although their masses decrease somewhat), continuously stirring
the disc and affecting the orbital decay of the SMBH binary. In most
cases, the dense gas clumps at early times convert most of their mass
into stars but remain gravitationally bound and only slowly get
disrupted. Note that due to the short timescales in question ($< 10$
Myr) to form the clumps and convert them into stars, the supernova
feedback has little effect on regulating the clump masses.

\begin{figure}
\centering
\includegraphics[width=\columnwidth]{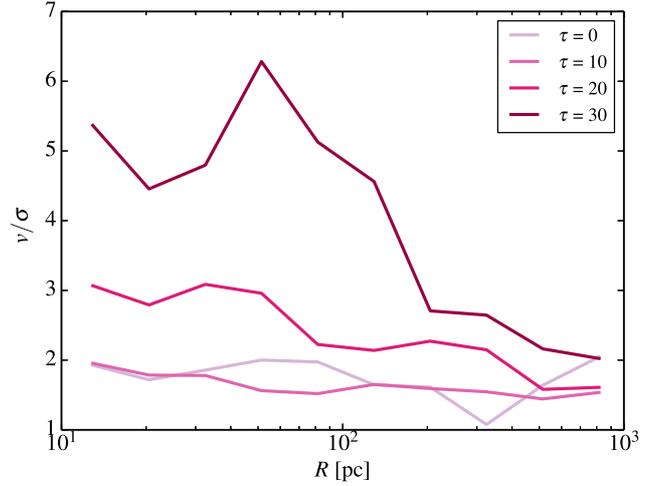}
\caption{The ratio of tangential velocity $v$ to velocity dispersion
  $\sigma$ at several times shortly after the merger is complete. The
  disc rebuilds on a timescale of $\sim 10^7$~years, evidenced by the
  rapid rise in $v/\sigma$, and extends out to $\sim 200$~pc.}
\label{fig:vsigma_short}
\end{figure}

Fig.~\ref{fig:vsigma_short} shows the ratio between rotational
velocity $v$ and velocity dispersion $\sigma$ of the gas disc,
quantitatively demonstrating the disc rebuilding phase. Immediately
after the merger ($\tau=0$ and $\tau=10$), $v/\sigma<2$, indicating
lack of rotational support and ordered bulk motion throughout the
central region. However, after just 4~Myr $v/\sigma > 2$ in the
interior 300~pc and by 10~Myr after the merger is complete, the inner
several hundred parsecs contain a rotationally-supported,
kinematically cold gas disc. Nevertheless, the gaseous disc remains
much less massive than the stellar component
(Fig.~\ref{fig:density_profiles}).

The gaseous and stellar discs are significantly flared, as shown by the
strong evolution of $z_{\rm rms}$ (model-independent proxy for scale
height) as a function of radius at all times. While the thickness of
the stellar component remains largely fixed throughout the 80 Myr of
evolution, however, the decrease in the thickness of the gas component
is evident especially in the inner regions. This is due to the
accumulation of gas in the central region as the disc reforms and the
chaotic merger-induced disc structure settles down. 

\begin{figure}
\centering
\includegraphics[width=\columnwidth]{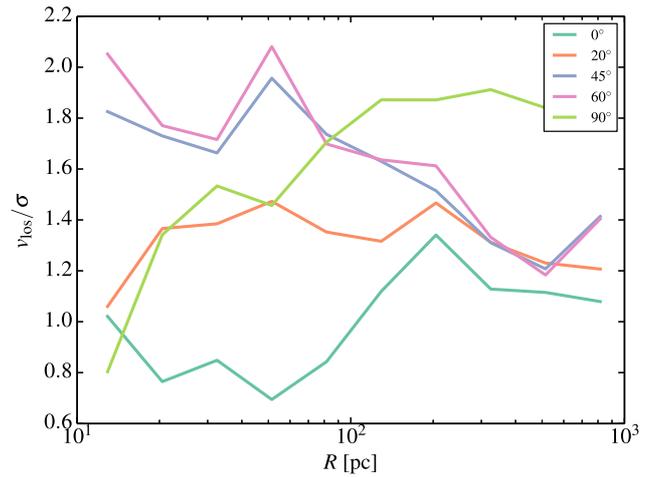}
\caption{$v_{\rm los}/\sigma$, where $v_{\rm los}$ is the mean velocity along
  the line of sight. Colours represent different inclinations of the
  central disc. The output shown is at $\tau = 30$ ($t = 5007$~Myr).
  }
\label{fig:vsigma_inclined}
\end{figure}

Fig.~\ref{fig:vsigma_short} shows the most ``optimistic'' measure of
$v/\sigma$, which is calculated simply as the ratio of the average
tangential velocity component and the velocity dispersion in each
radial bin. However, observationally, $v/\sigma$ is typically measured
from the line-of-sight velocity and the associated velocity dispersion
based on spectral line shift and width respectively. We crudely model
this type of $v/\sigma$ determination in
Fig.~\ref{fig:vsigma_inclined}, showing $v_{\rm los}/\sigma$ for a range
of inclination angles (0 corresponds to face-on) at $t = 5011
\mathrm{~Myr}$. The majority of $v_{\rm los}/\sigma$ values lie in the
range of 1-2, which signifies a thickened, highly turbulent disc.

In a recent study targetting 17 local (Ultra) Luminour Infrared
Galaxies (ULIRGs), \citet{Medling:2014} found that in most cases (16
out of 18 nuclei) they contain a dense nuclear disc composed of gas
and stars. The observed discs mostly have effective radii of a few
hundred parsecs and dynamical masses estimated in the range of $10^8 -
10^9 \mathrm{~M_{\odot}}$. The stellar populations appear young (ages
$< 10$~Myr), implying that they are associated with in-situ star
formation in the nuclear discs themselves, analogous to the disc in
our simulations. Furthermore, they find that $v/\sigma$ in their
sample ranges from 1-2, in good agreement with the values in
Fig.~\ref{fig:vsigma_inclined}. We can therefore be reasonably
confident that we are capturing some of the essential processes of the
nuclear disc rebuilding.

\subsection{SMBH Orbital Decay}
\label{sec:orbit_decay}

In the previous section we highlighted some of the most important
aspects of the morphological evolution of the merger remnant. The
morphology and the nature of the gas dynamics are critical for the
evolution of the SMBH binary. A prominent feature of the merger in our
simulation is that although in the initial conditions the system lies
in a single plane, this symmetry is broken during the final stages of
the merger (Fig.~\ref{fig:filmstrip}). As discussed above, this phase
is accompanied by the formation of massive clumps which additionally
perturb the SMBH orbits. The combination of these effects induces
significant vertical motion in the orbits of the two SMBHs with
respect to the the plane of the disc that is reforming after the
merger and it means that they both feel a much smaller drag on their
orbits around the centre compared to previous models
(e.g. \citealt{Mayer:2007}).

We have traced the evolution of massive clumps from one output to
another to determine their potential impact on the SMBHs. The time
between successive outputs is 50k years which means we can accurately
trace the orbits even in the innermost regions. We find that while
most clumps remain far enough that they do not have a significant
impact on the SMBH orbits, there is a small but important number of
clumps that come within 5-10 pc of the two SMBH particles. This
results in a random forcing on the orbits of the SMBHs which serves to
work against the drag from dynamical friction and helps keep one of
the SMBHs orbiting well out of the plane of the main disc. The ratio
of the SMBH velocity to the acceleration experienced due to an
individual clump, i.e. $v_{\rm bh}/a_{\rm clump}$ gives a timescale
for velocity change $\Delta v \propto v$. We find that this timescale
is less than a Myr for a number of clumps, suggesting that the random
forcing from clumps is significant for the orbital evolution of the
SMBH pair.

\begin{figure}
\centering
\includegraphics[width=\columnwidth]{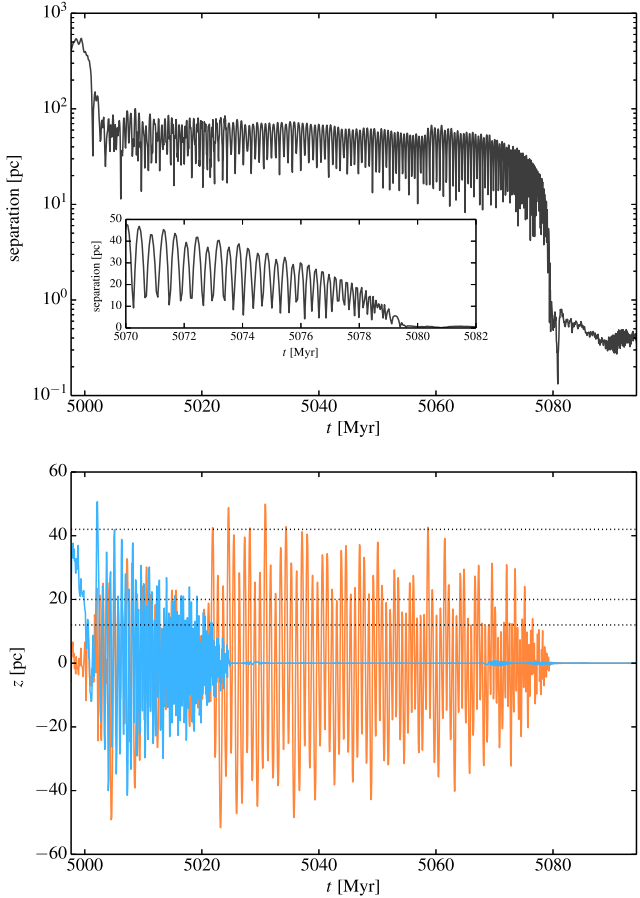}
\caption{{\bf Top}: Separation of the SMBHs as a function of time
  starting at approximately the last apocentre, i.e. the time when we
  switch to 1~pc spatial resolution. The inset shows the last stage of
  the decay. {\bf Bottom}: Motion perpendicular to the plane for each
  SMBH (shown with different colours). From top to bottom, the dotted
  horizontal lines show the $z_{\rm rms}$ of stars at 50, 10, and 5
  pc. }
\label{fig:r_z_vs_time}
\end{figure}

We show the orbital evolution of the SMBH binary after the final
apocentric passage in Fig.~\ref{fig:r_z_vs_time}. As described
above, the aftermath of the merger destroys the plane-symmetry of the
system and excites the out-of-plane oscillations of the two SMBHs (at
$\sim5002$~Myr). Because the densest component (stars and gas) is
distributed in a disc, the increased vertical oscillations result in
drastically reduced dynamical friction and therefore the SMBH orbital
decay timescale is extended. One of the SMBHs continues to decay its
orbit, albeit slowly, while the other remains at essentially a fixed
oscillation from the centre (Fig.~\ref{fig:r_z_vs_time}, bottom
panel). 

The decay (especially in the vertical oscillation amplitude) of the
binary continues until at $\sim5023$~Myr one of the SMBHs (blue)
begins to orbit within the misaligned nuclear disc.  As it nears the
centre, it encounters a weak stellar bar which measures a few parsec
across. The SMBH particle orbit is strongly torqued by the small bar
and it loses all of its angular momentum very rapidly, becoming pinned
to the potential minimum at the centre. The other SMBH particle
(orange) continues to experience random forcing from the clumps
(possibly also by spiral waves), and is therefore unable to sink to
the center until the clumps have largely dissolved several tens of Myr
later. We have estimated the dynamical friction timescale due to only
the background density (using calculations similar to those in
\citealt{Fiacconi:2013}) in the inner bulge to be $\sim 50$~Myr, so
the random forcings due to the clumpy medium increase this timescale
by a factor of $\sim2$. Note that during this time the SMBHs are
moving supersonically through the gaseous background at typical mach
numbers of 2-3. In principle this should amplify the drag
\citep{Ostriker:1999,Chapon:2013}, but the density of the gaseous
background is sufficiently low that the resulting dynamical friction
does not dominate the evolution.

The dotted lines in the bottom panel of Fig.~\ref{fig:r_z_vs_time}
show the vertical scale height of the stellar component at 5, 10 and
50 pc (bottom to top; see Fig.~\ref{fig:density_profiles}). During the
period that the decay of the second SMBH (orange line) seems stalled,
the vertical oscillations of the orbiting SMBH are beyond the vertical
scale height of the disc. Since much of the mass is concentrated in
the disc plane, these vertical excursions mean that the drag on the
SMBH orbit is reduced. Furthermore, dynamical friction against the gas
is more efficient where the bulk motion is highly ordered, which is
not the case outside the disc. The final decay is shown on the inset
axes in the top panel.

During the last few Myr of the decay, the orbit of the second SMBH is
eccentric with $e = (r_a - r_p)/(r_a + r_p) = 0.7$ and just before the
binary decays to our resolution limit $e \sim 0.4$. This is consistent
with \citet{Fiacconi:2013}, where it was found that for in the SMBH
decay with a clumpy ISM, the orbits remained eccentric down to the
resolution limit. The fact that the binary approaches the
gravitational wave regime before circularizing can be critical for the
subsequent gravitational wave emission \citep{Mayer:2013}.

Finally, an important caveat in the evolution of the SMBH binary is
the fact that our simulation does not include the effects of BH
accretion and associated AGN feedback. We perform a crude estimate of
the expected accretion rate assuming spherical accretion and taking
into account the relative velocity between the black hole and the
background
\begin{equation}
\dot{M} = \frac{4 \pi G^2 M_{\bullet}^2 \rho}{(v_{\mathrm{rel}}^2 + c_{\mathrm{s,turb}}^2)^{3/2}},
\label{eq:bondi}
\end{equation}
where $M_{\bullet}$ is the mass of the SMBH, $v_{\mathrm{rel}}$ is the
relative velocity between the SMBH and the background,
$c_{\mathrm{s,turb}} = (c_s^2 + \sigma^2/3)^{1/2}$ is the turbulent
sound speed with $\sigma$ the gas velocity dispersion and $G$ is the
gravitational constant. Insofar as the SMBH particles can be
considered as test particles sampling the potential in the inner part
of the galaxy, we approximate $v_{\mathrm{rel}}$ as the velocity
dispersion of the stellar component which during the time shortly
after the merger is $\sim 200$~km/s. The resulting Bondi rates are
$\sim0.05-0.1$~M$_{\odot}$ yr$^{-1}$. This means the BHs could double
their mass during the $\sim 100$ Myr of decay phase in the nuclear
disk, which would have only a marginal effect on dynamical friction as
well as on the exchange of energy and angular momentum via stochastic
torquing.  Using these approximations and assuming that the luminosity
of the nuclei is given by $L = \epsilon_r \dot{M} c^2$ where
$\epsilon_r$ is $\sim 10\%$ from both observational constraints and
models of viscous accretion disks.  With these assumptions, we
estimate that, for the typical gas densities encountered by the SMBHs
(recall that the SMBHs oscillate significantly above and below the
disk plane) the expected luminosities would be 
$< 0.1 L_{\mathrm{edd}}$, i.e. $\sim 10^{43}$ erg~s$^{-1}$.  Of this
accretion luminosity only a very small fraction would couple to the
ISM, thermally and/or via momentum transfer.  While the coupling
mechanism and efficiency is largely unknown, simulations that attempt
to reproduce the observed correlations between SMBH masses and host
galaxy properties require a coupling efficiency $\epsilon_{fb} \sim
0.005-0.05$, where the wide range of values is explained by the
dependence on the specific thermodynamical model of the ISM that
simulations adopt (see e.g. \citealt{Springel:2005,Callegari:2009}).
Therefore, we can assume the luminosity actually coupled with the gas
will be  $< 10^{42}$ erg/s.  We can compare this upper limit
with the energy actually released into the ISM by supernovae type II
via our blastwave feedback recipe during the phase of SMBH decay in
the nuclear disk.  This amounts to $\sim 10^{42}$ erg/s. This is a
conservative estimate since it does not include the energy released by
SN type I, which we include as thermal energy injection into the ISM
in the simulations but allow it to be radiated away rather than
modeling it with a blastwave (see \citealt{Stinson:2006}). Overall,
these calculations suggest that the impact of AGN feedback in the SMBH
decay phase studied in this paper would be modest since it will likely
deposit less energy then supernovae feedback. Of course there are huge
uncertainties in how to model the various feedback processes in the
first place. However the main point here is that we are already
depositing a significant amount of thermal energy into the ISM with
our feedback model, hence the existence of a prominent cold clumpy
phase in the nuclear disk should be viewed as a fairly robust outcome.

\section{Conclusions}
\label{sec:conclusions}

We perform a multi-scale simulation of an equal-mass merger of two
spiral galaxies with an SMBH embedded in each of the progenitor's
centers. As the merger nears completion, we increase the resolution of
the central region via a particle splitting method, taking care to
ensure that the thermodynamic state of the gas and the instantaneous
properties of the feedback processes are preserved. Throughout the
simulation, our model includes star formation and stellar feedback,
establishing a multiphase, turbulent ISM in the central region. After
the two cores successfully merge and the effects of the nuclear
starburst subside, a nuclear disc grows in the centre of the merger
remnant in the span of a few Myr, eventually reaching a size of
$\sim400$~pc. The two SMBHs form a loose binary at this point, but the
violent nature of the late stages of the merger causes their orbits to
have a considerable motion perpendicular to the disc plane. This
delays the eventual complete decay of the SMBH binary which takes
approximately 80~Myr.  This decay timescale is comparable to that
found in recent idealized experiments (supplemented by analytical
estimates) by \citet{Fiacconi:2013} and almost two orders of magnitude
longer than found in previous studies of analogous systems that used
an effective equation of state to model the gas
\citep{Mayer:2007}. 

Our results indicate that SMBH pairs may decay to parsec scales on a
much longer timescale than previously thought.
If we consider the timescale $\sim100$~Myr to form a tight binary from
a separation $\sim100$~pc as typical, galaxies that show morphological
signatures of a recent merger (e.g. tidal tails) with a post-starburst
stellar population might be the most appealing candidates to look for
SMBH pairs.  However, this timescale is subordinate to the dynamics of
the SMBH pair in the inhomogeneous background and to the eventual
presence of massive enough clumps formed during the merger.  Moreover,
it is worth recalling that the stochastic character of the dynamics
outlined by our results may also accelerate the orbital decay process.
\citet{Fiacconi:2013} estimated a threshold mass
$\mathcal{M}_{\bullet}$ below which a SMBH orbiting in such an
environment might be dynamically influenced by massive clumps.
Although it was only an order of magnitude estimate, a typical value
$\mathcal{M}_{\bullet} \sim 10^{7}$ M$_{\odot}$ is fairly consistent
with our results and with a recent discovery of SMBH pair by
\citet{Fabbiano:2011}.  This mass roughly corresponds to a bulge mass
$\lesssim 5 \times 10^{9}$ M$_{\odot}$, according to the scaling
relation between SMBH and bulge masses \citep{Sani:2011}.  This
suggests an additional conservative selection criterion: SMBH pairs
might be more easily detected in late-type (hence relatively gas-rich
even at $z=0$) galaxies with stellar mass $\lesssim
10^{10}$~M$_{\odot}$, although the occupation fraction of SMBHs at
such low masses is still unclear. 

Nevertheless, these considerations still neglect the details of the
dual AGN activity required for the SMBHs to be detected
\citep{vanWassenhove:2012,vanWassenhove:2014}.  Just after the merger
phase, obscuration might be relevant or the central gas might be
depleted after the starburst and mostly accumulated in to massive
clumps, reducing the time during which the SMBHs could be active at
the same time. Our simulations here do not include a model for SMBH
accretion and AGN feedback, which would be critical for determining
reasonable observability constraints.

The simulations presented here used galaxy models that were tailored
to reproduce low-redshift late-type galaxies. This was motivated
primarily by the need to compare with previous work using similar
initial conditions (e.g. \citealt{Mayer:2007}). Furthermore such
systems host moderate mass SMBHs, which would be the preferential
target of future gravitational wave experiments such as eLISA making
our choices of broader relevance. However, since galaxy mergers are
much more frequent at higher redshift, future work will have to
explore mergers between galaxies whose properties are more akin to
high-z galaxies. The latter will be needed to assess the likelihood of
finding SMBH binaries at tens of parsecs to a few parsecs separations
as well as being more relevant to make predictions for rates of
gravitational wave emission events originating from coalescing massive
BH binaries. Qualitatively, we can expect that the stochastic orbital
decay regime found here will be even more relevant because at $z>1$
galaxies with stellar masses comparable to the Milky Way appear to be
more gas-rich as well as clumpier even when they are not in a merging
phase \citep{Genzel:2006, Elmegreen:2010}. Clumps observed
in high-z galaxies are also much more massive, up to a few $10^8$ 
M$_{\odot}$, which would imply stronger gravitational forcing of the
massive BHs during encounters, potentially leading to even larger
delays in the orbital decay. As shown by \citet{Fiacconi:2013}, the
larger mass scale of the clumps also means stochastic torques will
affect comparatively more massive BHs, in the range $10^7-10^8$
M$_{\odot}$. Such gas-rich, clumpy massive disc galaxies at $z > 2$ with
relatively massive central SMBHs are likely the progenitors of massive
elliptical/S0 galaxies today, hence the stochastic torquing regime may
be particularly relevant to understand the dynamical evolution and
mass growth of the most massive SMBH found in the current
Universe. 

While quantitative statements cannot be made at this stage, it is
rather plausible that dual AGNs with separations of tens of parsecs
should be quite common at $z > 1$ due to combination of long orbital
decay timescales implied by a clumpy ISM, perhaps exceeding 100 Myr,
and the large gas reservoirs as well as highly dynamical environments
leading to efficient gas inflows and accretion onto SMBHs
(e.g. \citealt{Bournaud:2012}). The dense gaseous environments
expected at such epochs may lead to widespread obscuration of the AGN
emission at UV and optical wavelengths, rendering the detection by
ALMA and other long wavelength instruments difficult.  Therefore
hi-resolution X-ray space observatories, such as ATHENA, will probably
be needed to really assess the abundance of such small separation dual
AGNs at high-z.

\section*{Acknowledgments}
The authors wish to thank Marco Spaans for providing the data tables
used for the equilibrium temperature corrections.  RR is funded in
part by a Marie Curie Career Integration Grant. Part of this research
was funded by NASA Award NNX07AH03G. RR also gratefully acknowledges
the Aspen Center for Physics, funded by the NSF Grant \#1066293, for
hospitality during the writing of this manuscript.

\bibliographystyle{mn2e} 
\bibliography{smbh_p1}

\end{document}